\documentclass[prb,aps,showpacs,twocolumn,floatfix]{revtex4}
\usepackage{graphicx}
\begin{document}
\title{Nanotube Piezoelectricity}
\author {Na Sai and E.J. Mele}
\affiliation{Department of Physics and Astronomy, University of Pennsylvania,
Philadelphia, PA 19104}
\date{August 21, 2003}
\begin{abstract} 
We combine {\it ab initio}, tight-binding methods and analytical theory 
to study piezoelectric effect of boron nitride
nanotubes. We find that piezoelectricity of a heteropolar nanotube
depends on its chirality and diameter and can be understood starting from the
piezoelectric response of an isolated planar sheet, along with a
structure specific mapping from the sheet onto the tube
surface. We demonstrate that coupling between the uniaxial and shear 
deformation are only allowed in the nanotubes with lower chiral
symmetry. Our study shows that piezoelectricity of nanotubes is
fundamentally different from its counterpart in three dimensional (3D)
bulk materials.   
\end{abstract}
\maketitle
The physical properties of a nanotube along its extended
direction are controlled by the boundary conditions imposed along its
wrapped direction.  The existence of both semiconducting and metallic
forms of pure carbon nanotubes provides a striking example.\cite{book}
The recently discovered electric polarization in heteropolar nanotubes
(e.g. Boron-Nitride) presents a new physical manifestation of this
effect.\cite{Mele} Since the polarization can be modulated by elastic
strains of the tube, these materials provide a new class of {\it
molecular piezoelectrics} where mechanical strain is linearly coupled to
an electric field. Piezoelectric nanotubes thus hold promise for
application in nanometer scale sensors and actuators. 

In the modern quantum theory of polarized solids the electric
polarization is computed from the geometric phase \cite{King-Smith}
(Berry's phase) accumulated by the occupied electronic states as one
introduces a potential that adiabatically connects an unpolarized and
polarized state of the system. For a BN nanotube the Berry's phase and
hence the polarization is controlled by the periodic boundary condition
on electronic wavefunctions.\cite{Mele} Piezoelectric effect, on the
other hand, is determined by the dependence of macroscopic polarization
on the local strain induced effects:  redistribution of the valence
charge density, curvature induced rehybridization of the electronic
orbitals and relaxations of the positions of the atoms on the tube walls
which are all short range in character. Here we show that it is this
latter character that allows the piezoelectric response to follow a
simple transformation rule when the structure changes from a sheet to
tube geometry.

\begin{figure}
\includegraphics[width=6.0cm, angle=0]{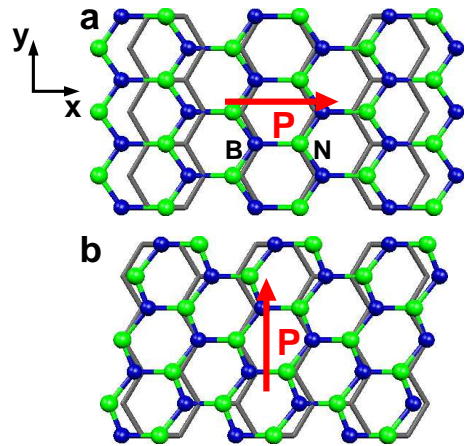}
\caption{BN flat sheets under uniaxial strain $\eta_{xx}$ (a) and shear strain
$\eta_{xy}$ (b).
In both cases, threefold symmetry is broken and charge redistribution
gives rise to a net dipole moment. The corresponding polarization 
directions ($P$) are marked by arrows.}
\label{fig:2d}
\end{figure}

The prototypical example of piezoelectric nanotubes is found in the
family of BN nanotubes where the alternation of group III(B) and group
V(N) elements on the honeycomb lattice lowers the symmetry. A BN nanotube
can have a nonzero electric polarization \cite{ionic} unlike its planar
counterpart where this is forbidden by the threefold rotational symmetry
of an isolated two dimensional (2D) BN sheet. However an elastic coplanar deformation of a BN
sheet lowers its lattice symmetry, redistributes the valence charge and
produces a nonzero polarization. Fig.~\ref{fig:2d} illustrates the effect
of a uniaxial strain ($\eta_{xx}$) and a shear strain ($\eta_{yx}$) of
the BN sheet, with both distortions greatly exaggerated for clarity.
These distortions induce the electric dipole moments denoted by the
arrows. The linear response of the electric polarization $P_i$ to an
applied strain $\eta_{jk}$ is described by the third rank piezoelectric
tensor $e_{ijk} = \partial P_i/\partial \eta_{jk}$. The $3m$ symmetry of
the unstrained sheet requires that the piezoelectric tensor is unchanged
by threefold rotations of the lattice and the elements of the
piezoelectric tensor obey the symmetry relation
$e_{xxx}=-e_{xyy}=-e_{yxy}=-e_{yyx}$.

To study the microscopic origin of this behavior we first carried out
{\it ab initio} calculations of the piezoelectric constants of the flat
BN sheet using a planewave pseudopotential method based on
density-functional theory within the local density approximation. The
calculation is performed with the {\tt Abinit} package \cite{Abinit}
using Troullier-Martins pseudopotentials \cite{TM} with an energy cutoff
of 45 Hartree and $4\times4\times1$ $k$-point grid throughout. To create
a computational cell that is periodic in all three spatial dimensions we
stacked the BN sheets with an interplanar distance 20 Bohr so that there
is negligible wave function overlap between layers. The electronic
polarization was computed for a series of strained lattices using the
Berry's phase formulation \cite{King-Smith} (discretized on a dense
$k$-point grid along the direction of the polarization) and the
piezoelectric constants were obtained by calculating lattices with
strains in the range of $1\%\le\eta_{jk}\le5\%$. By defining the positive
direction to be the bond direction from B atom to N atom as shown in
Fig.~\ref{fig:2d}, we find $e_{xxx}=-0.12 \, {\rm e/Bohr}$.  Our
calculations for the sheets with a shear strain $\eta_{xy}$ have
explicitly verified the symmetry relations among the piezoelectric tensor
elements and showed that the piezoelectric properties of the sheet are
controlled by a single coefficient. To carry out a systematic study of
the piezoelectric behavior of a large family of wrapped structures
parameterized by integer indices $(m,n)$, \cite{book} we combine the {\it
ab initio} DFT method with a computationally less intensive albeit less
accurate tight-binding (TB) method. We used a non-orthogonal basis set
with four orbitals per site to describe the $2s$ and $2p$ atomic
orbitals. \cite{Porezag} We were able to benchmark our tight-binding
method by comparing calculations of the piezoelectric constant of the BN
sheet using both theories. We find that the TB theory yields $e_{xxx}
=-0.086 \, {\rm e/Bohr}$ which is smaller than the {\it ab initio}
result, though in acceptable agreement.

Note that the value of piezoelectric constant of a 2D BN sheet has a dimension charge per unit length. A quantitative
comparison of the piezoelectric constant of the flat sheet to the
piezoelectric coefficients of 3D bulk material requires specification of
the interlayer spacing and packing. For example, if we convert the 
above 2D flat sheet value into a conventional ``bulk'' piezoelectric
constant using the primitive interlayer separation of $0.34 \, {\rm
  nm}$,\cite{Lu} we find $e_{3D} = 0.76\, {\rm C/m^2}$. This value is similar
in magnitude to $e_{33} = 0.73\, {\rm C/m^2}$ of wultzite nitrides (e.g.
GaN)\cite{Bernardini} and is larger than $0.12\,{\rm C/m^2}$ of piezoelectric polymer polyvinylidene
fluoride(PVDF-TFE).\cite{Kocher} Alternatively the piezoelectric constant
computed from different dimensional system can be expressed as total
dipole per stoichiometric unit. Using this convention, we find that the
piezoelectric constant is 1.67 dipole/unit for the BN sheet, smaller than
1.98 dipole/unit for GaN wultzite.  As we show below, when a sheet is
wrapped to form a tube, there is a unique well defined relation between
the piezoelectric constants of a tube and flat sheet.

\begin{figure}
\includegraphics[width = 9.0cm, angle=0]{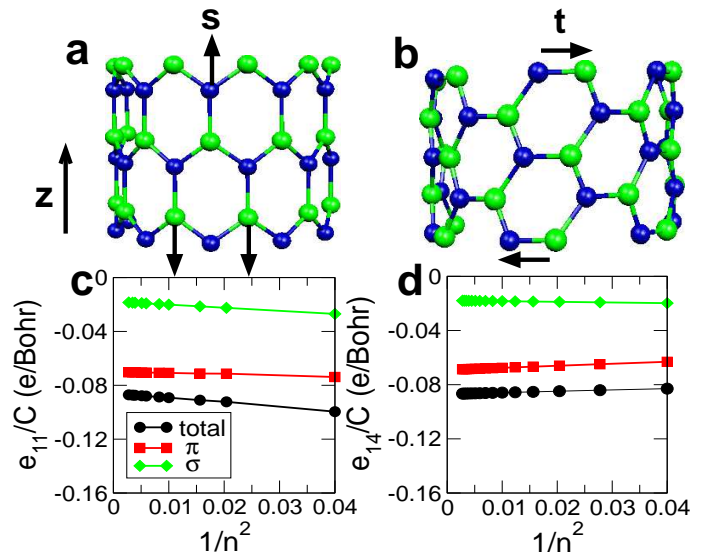}
\caption{Schematic structures of (a) stretched
$(n,0)$ nanotubes and (b) twisted $(n,n)$ nanotubes where arrows display
the strain deformation direction in the tangential plane. Panels (c) and
(d) show the calculated tube piezoelectric constant $e_{11}/C$
and $e_{14}/C$ as functions of $1/n^2$ where $C$ is the tube
circumference. Contributions from
$\sigma$ and $\pi$ electrons to the total piezoelectric response are separated.}
\label{fig:zig_arm}
\end{figure}
Two high symmetry families of nanotubes are the zig-zag structures with
wrapping indices $(n,0)$ and the armchair structures with wrapping
indices $(n,n)$. The one dimensional(1D) piezoelectric constants are defined
as $e_{11} =\partial P_z/\partial \eta_s$ and $e_{14} = \partial
P_z/\partial \eta_t$, where $P_z$ is the dipole moment per unit length
and the $(z,s,t)$ indices in the tube frame refer, respectively, to the
tube axis ($z$) and the uniaxial ($s$) and torsional ($t$) strains. In
Fig.~\ref{fig:zig_arm}, the top panel shows the structures of two
representative small radius members of each family and in the bottom
panel we plot their piezoelectric constants in unit comparable with the
two dimensional piezoelectric constants. Note that the 1D
piezoelectric constant is proportional to the tube circumference $C$
through $e_{1D} = C\,e_{2D}$. We find that zigzag tubes exhibit a
longitudinal piezoelectric response for the case of uniaxial strain
(extension or compression) but not for torsion. In contrast the armchair
tubes have an electric dipole moment linearly coupled to torsion, but not
to a uniaxial strain. The complementary strains, i.e. torsion for the
zigzag structures and stretch for the armchair structures, produce a
purely azimuthal dipole that integrates to zero on the surface of the
cylinder.

For large radius tubes, one expects a correction to the piezoelectric
constants of tube from its curvature, proportional to the inverse square
of the tube radius. This can be seen in Fig.~\ref{fig:zig_arm} bottom
panel where we quantify this scaling behavior by plotting the calculated
piezoelectric constants as a function of $1/n^2$. The data show that the
tube piezoelectric constant rapidly approach the flat sheet values with
this scaling relation, but also that curvature effects remain quite small
even for relatively small radius tubes. The data also show that the $\pi$
and $\sigma$ valence electrons have the same sign for both families of
structures, with the $\pi$ electrons dominating the piezoelectric
response, accounting for approximately $80 \%$ of the total.

\begin{figure}
\includegraphics[width=9.0cm, angle=0]{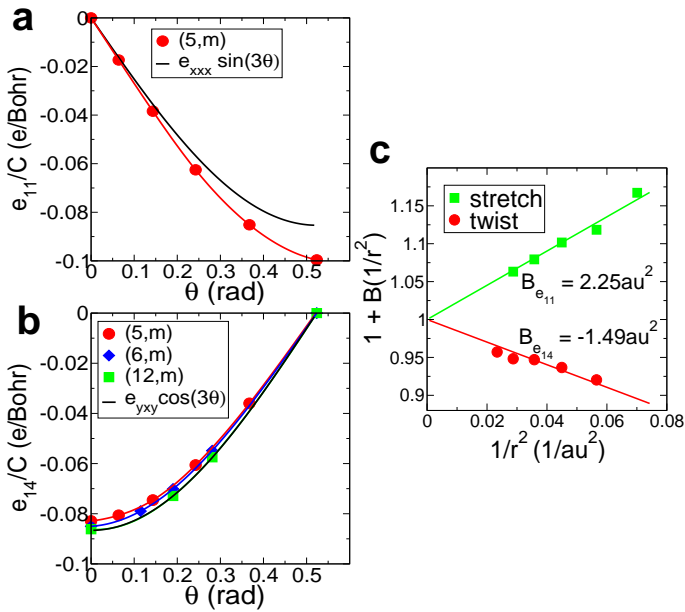}
\caption{Piezoelectric response as function of the chiral
angle in a sample of chiral nanotubes experiencing
the uniaxial strain (a) and the shear strain (b). Solid black curves
are the analytical result, Equation~(\ref{eq:e}).
Panel {\bf c} shows ratio of the piezoelectric constants of chiral nanotubes
to their flat sheet values plotted as a function of the inverse square
radius. The two branches are for the uniaxial $e_{\rm 11}$ and torsional
$e_{\rm 14}$ response.}
\label{fig:chiral}
\end{figure}
For a chiral tube the wrapping vector does not lie along a high symmetry
direction of the 2D honeycomb lattice. This leads to a large and low
symmetry translational unit cell for the chiral nanotube making a direct
calculation of its piezoelectric properties cumbersome. We make use of
the results for the high symmetry armchair and zigzag structures to
develop an accurate scaling theory of the piezoelectric response of
chiral tubes. Ignoring the finite radius corrections arising from the
tube curvature the elements of the piezoelectric tensor are specified by
rotating the known piezoelectric elements of the flat sheet onto the
symmetry axes of the tube. Thus, defining the chiral angle $\theta$ as
the angle between the axis of the tube and a 2D primitive translation
vector we find 
\begin{eqnarray}
e_{11} &=& C\,e_{xxx}\sin(3\theta) \nonumber\\
e_{14} &=& C\,e_{yxy}\cos(3\theta).
\label{eq:e}
\end{eqnarray}
Finite radius corrections to the predictions of Equation (\ref{eq:e}) can
then be obtained by comparing the results of this mapping to the values
obtained from TB calculations on a selected set of chiral structures.
Fig.~\ref{fig:chiral}a and b shows the result of this comparison, for
uniaxial strain on the family of $(5,m)$ tubes and for torsion on the
families of $(5,m)$, $(6,m)$ and $(12,m)$ families respectively. These
data are very well described by the mapping of the 2D results, with
surprisingly small corrections due to the tube curvature. The correction
is quantified in Fig.~\ref{fig:chiral}c where we plot the ratio of the
calculated piezoelectric modulus to its flat sheet ($n \rightarrow
\infty$) value as a function of $1/r^2$ for the family of $(5,m)$ tubes.
The deviations from the predictions of the flat sheet model are less than
$15\%$ over the entire range of structures we studied.

Equation (\ref{eq:e}) also reveals that the chiral tube has allowed
linear piezoelectric coupling to both uniaxial strain {\it and} torsion,
unlike the higher symmetry zigzag or armchair structures. Thus the long
wavelength elastic energy of a chiral tube generically has an anomalous
cross term containing the product of the uniaxial and torsional strains.
This implies that a tensile stress applied to a chiral tube induces
torsion and conversely torsion induces a change of its length. Such a
coupling is only possible for a chiral molecular structure, and indeed
the coefficient of the cross term is a macroscopic manifestation of the
underlying microscopic chirality of the nanotube.

Recent progress in the synthesis of nanoscale materials  is
demonstrating that many three dimensional lamellar phases can be
fabricated in compact cylindrical structures.\cite{Tenner} The appearance of
pyroelectric and piezoelectric effects is a generic feature of these
structures, and can be excluded only for special high symmetry wrappings.
The methods we have developed and tested here for BN nanotubes should be
widely applicable to study piezoelectric effects in this broader family
of nanoscale materials.

\begin{acknowledgments} 
This work was supported by Department of Energy
under grant DE-FG02-ER0145118 and by National Science Foundation under
grant DMR-00-79909. We thank J. Bernholc, S. Nakhmanson and P. Lammert
for helpful discussions.  
\end{acknowledgments}

\end{document}